\documentclass[conference]{IEEEtran}
\IEEEoverridecommandlockouts
\usepackage{cite}
\usepackage{amsmath,amssymb,amsfonts}
\usepackage{algorithmic}
\usepackage{graphicx}
\usepackage{makecell}
\usepackage{textcomp}
\usepackage{float}
\usepackage{color}
\usepackage{tikz}
\usepackage{pgfplots}
\usepackage{xcolor}
\usepackage{comment}
\usepackage{booktabs}
\usepackage{siunitx}
\usepackage{tabu}
\usepgfplotslibrary{statistics}
\usepackage[
top=50pt,
bottom=54pt,
left=46pt,
right=46pt
]{geometry}
\pgfplotsset{compat=1.18}
\def\BibTeX{{\rm B\kern-.05em{\sc i\kern-.025em b}\kern-.08em
    T\kern-.1667em\lower.7ex\hbox{E}\kern-.125emX}}
\begin{document}

\IEEEaftertitletext{\vspace{-2.5\baselineskip}}

\title{
\vspace{15px}Performance Characterization of Frequency-Selective Wireless Power Transfer Toward Scalable Untethered Magnetic Actuation\\
\thanks{\hspace*{-1em}*Corresponding author: Xiaolong Liu (email: xiaolong.liu@ttu.edu).\\
This work was supported in part by the National Science Foundation under Grant No. 2512394, in part by the Texas Tech HEF New Faculty Startup Fund, and in part by the Texas Tech TrUE and McNair Scholar Programs.}
}

\author{\IEEEauthorblockN{Gabriel Cooper}
\IEEEauthorblockA{\textit{Department of Electrical and Computer Engineering} \\
\textit{Texas Tech University}\\
Lubbock, USA\\
}
\and
\IEEEauthorblockN{Xiaolong Liu$^*$}
\IEEEauthorblockA{\textit{Department of Mechanical Engineering} \\
\textit{Texas Tech University}\\
Lubbock, USA \\
}
\vspace*{-100px}
}

\maketitle

\begin{abstract}
Frequency-selective wireless power transfer provides a feasible route to enable independent actuation and control of multiple untethered robots in a common workspace; however, the scalability remains unquantified, particularly the maximum number of resonators that can be reliably addressed within a given frequency bandwidth. To address this, we formulate the relationship between resonator quality factor (Q-factor) and the number of individually addressable inductor-capacitor (LC) resonant energy harvesters within a fixed radio-frequency (RF) spectrum, and we convert selectively activated harvested energy into mechanical motion. We theoretically proved and experimentally demonstrated that scalability depends primarily on the Q-factor. For this proof-of-concept study, we define effective series resistance as a function of frequency allocating bandwidths to discrete actuators. We provide design equations for scaling untethered magnetic actuation with Q-factor optimization. Resonator networks spanning bandwidths from 100~kHz to 1~MHz were analyzed to quantify how increasing the number of resonators affects independent addressability. We validated the approach experimentally by fabricating three centimeter-scale untethered actuators that selectively trigger the motion of mechanical beams at 734~kHz, 785~kHz, and 855~kHz. We also characterized the generated mechanical force and the activation bandwidth of each actuator, confirming that no unintended cross-triggering occurred.
\end{abstract}


\section{Introduction}
\IEEEPARstart{S}{elective} actuation of untethered magnetic robots operating in a shared workspace is a fundamental challenge because their actuation is inherently coupled through a common magnetic field source. Prior work has explored two main strategies for achieving selective actuation of magnetic robots, including the use of inhomogeneous magnetic fields to actuate robots at different spatial locations~\cite{7394119, 8440784,525362,Chowdhury_Jing_Cappelleri_2015, RichterAdvs2023LocallyAddressable} and rotational step-out behavior in robots with different geometries or configurations~\cite{Floyd_Diller_Pawashe_Sitti_2011, 6056575}. For example, global magnetic field dynamics exploit variations in magnetic material composition~\cite{Floyd_Diller_Pawashe_Sitti_2011}, robot geometry~\cite{6056575,Tottori_Zhang_Peyer_Nelson_2013}, or robot placement in field gradients~\cite{7394119} to produce differentiated responses under a shared input field. Alternative approaches have achieved localized magnetic manipulation using two-dimensional (2-D) electromagnetic coil arrays~\cite{8440784,525362,Chowdhury_Jing_Cappelleri_2015}, which generate spatially controllable fields in 2-D spaces and enable independent control based on robot position. Rotational step-out frequency based actuation has also been investigated using rotating magnetic fields to drive robots with different geometries that experience different fluid drag, including helical swimmers~\cite{Amoudruz_Koumoutsakos_2021,Ishiyama_Sendoh_Arai_2002}, and block-shaped robots~\cite{6056575}. These approaches provide important starting points, but they remain difficult to scale and generalize.

Beyond approaches that rely on low frequency magnetic fields to achieve selective actuation of magnetic robots, wireless power transfer (WPT) techniques have been used to deliver RF magnetic energy to resonant circuits integrated into robotic devices, enabling actuation through mechanisms such as magnetic propulsion~\cite{8361455}, heating shape memory alloy folding actuators~\cite{Boyvat_Koh_Wood_2017}, frequency-selective heating of soft actuators using embedded resonant traces~\cite{SongNatCommun2025RFSelectiveLCE}, selective drive of electrostatic actuators by tuning the transmitter frequency to the receiver LC resonance~\cite{TakeuchiSensActA2002SelectiveDrive}, or wireless actuation of micron-scale piezoelectric resonators at $36-120$~\si{MHz} using RF electric-field coupling~\cite{MateenMicrosystNanoeng2016WirelessActuation}. Despite these advances, selectively actuating multiple untethered magnetic robots using RF magnetic energy is challenging due to field coupling, limited spectral separation between robot responses, and the increasing risk of cross actuation as the robot population grows. Additionally, at small scales, robots have limitations for onboard energy storage and power delivery. Actuation requires sufficient harvested energy to perform meaningful work, yet the size constraints of miniature robots restrict the inclusion of large batteries, power electronics, and control systems. As a result, scaling selective actuation to a large number of small-scale robots remains an open problem. In particular, the relationship between Q-factor and the number of individually addressable untethered robots within a shared space has not been systematically studied.

In this work, we quantify the scalability of frequency-selective WPT for untethered robotic actuation by deriving and validating how resonator Q-factor sets the maximum number of individually addressable LC resonant harvesters within a fixed RF bandwidth, and by demonstrating selective conversion of harvested energy into mechanical motion. Our approach integrates frequency-selective LC resonators with RF energy harvesting circuits to convert received RF power into on-demand magnetic actuation. We focus on the \SI{100}{kHz} to \SI{1}{MHz} band because it offers improved penetration through surrounding media and reduced eddy current losses relative to higher frequency operation. This lower frequency range also simplifies circuit design by reducing the need for transmission line modeling.

To study scalability, we develop a design equation relating resonator losses, circuit topology, and Q-factor to the achievable frequency bandwidth for selective actuation. This model provides insight into the maximum number of robots that can be independently addressed within a given frequency spectrum. We experimentally validated the model using centimeter-scale robotic devices containing LC resonator networks designed for energy harvesting and electromagnetic (EM) pulse generation. Three devices were fabricated, each with two LC resonators (one for charging and one for selective triggering), on printed circuit boards (PCBs). The resonators were configured for rapid energy accumulation and triggered the EM pulse actuation of hard-magnetic cantilever beams. Experimental measurements quantify activation bandwidth, triggering selectivity, and force output, and they directly validate the predicted scaling trends from the design equation.

The main contributions of this work are as follows: 
\begin{enumerate}
\item We develop a theoretical framework that relates resonator Q-factor, circuit losses, and topology to the scalability of RF selective actuation for untethered robots.
\item We validate RF energy-harvesting-driven magnetic actuation using integrated LC resonator networks.
\item We demonstrate frequency-selective actuation of three individually addressable untethered robots via resonator-triggered EM pulse actuation.
\item We provide design guidelines for scaling RF-driven magnetic robots toward smaller devices and larger robot populations.
\end{enumerate}

\section{RF Selective Actuation Mechanism}
\subsection{Working Principle}

\begin{figure}[t]
    \centering
    \vspace{4pt}
    \includegraphics[width = 1\linewidth]{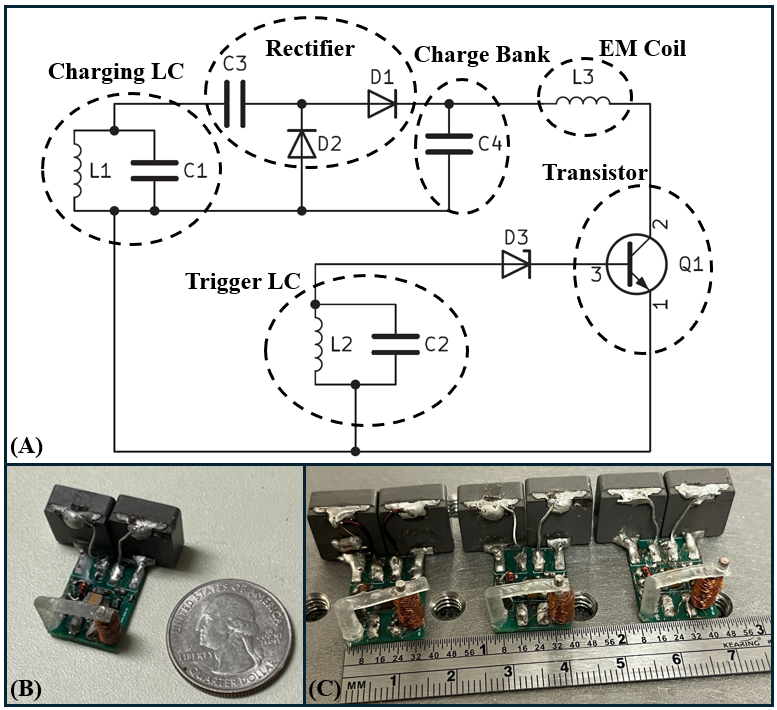}
    \caption{Working principle of the RF selective actuation mechanism and prototype demonstration. (A) Schematic diagram of the circuit design. (B) Prototype next to a quarter for size reference. (C) Three prototypes with different triggering frequencies.}
    \label{fig:1}
    \vspace{-15px}
\end{figure}

Fig.\ref{fig:1}(a) illustrates the working principle of the proposed RF selective actuation mechanism. Each unit includes an energy-harvesting LC circuit (L1, C1) tuned to a common resonant frequency 
\begin{equation}\label{eq:1}
    f_0 = \frac{1}{2\pi\sqrt{LC}},
\end{equation}
and the harvested energy is stored in a power bank capacitor (C4) through a rectifier stage (C3, D2, D1). Each unit also incorporates a secondary LC circuit (L2, C2) with a distinct resonant frequency that selectively drives the gate of an NPN transistor Q1, allowing the stored energy to discharge through an onboard EM coil (L3). The resulting coil current generates a magnetic interaction with an integrated mechanical structure to produce actuation. Fig.\ref{fig:1}(b) shows a fabricated single robotic unit integrating the circuit and mechanical structure, and Fig.~\ref{fig:1}(c) shows three identical devices with different triggering frequencies.

The energy stored in capacitor C4 is determined by its voltage V as
\begin{equation}\label{eq:2}
E = \frac{1}{2}CV^2.
    \vspace{2px}
\end{equation}
As the mutual inductance between the transmitter ($L_T$) and the receivers ($L_R$) is defined by
\begin{equation}\label{eq:3}
    M = k\sqrt{L_TL_R}
\end{equation}
where the coupling coefficient $k$ is primarily determined by the coil geometry and relative placement, as well as the materials and electromagnetic environment of the surrounding medium. 

For experimental convenience to validate our system, the initial surface-mount inductors were swapped with larger inductors, as shown in Fig.~\ref{fig:1}(b), to increase mutual inductance and the coupling coefficient to provide sufficient energy to saturate the transistor. To significantly reduce the design footprint in our future work, we will employ a MOSFET with a high input impedance gate to minimize steady state drive current and reduce loading on the trigger LC resonator. This helps maintain a high loaded Q-factor and keeps more energy in the resonator for triggering, which is important as the device is miniaturized and WPT becomes less efficient.

\subsection{Role of Q-factor}\label{subsec:q_factor}
Q-factor determines how sharp each trigger resonator's frequency response is. For a lightly damped LC resonator, the half-power bandwidth, also referred to as the \SI{-3}{\decibel} bandwidth, $\Delta f$ satisfies
\begin{equation}\label{eq:4}
\Delta f \approx \frac{f_0}{Q},
\end{equation}
where $f_0$ is the resonant frequency of the LC circuit~\cite{Coakley2003QEstimation}. A smaller $\Delta f$ corresponds to a larger Q-factor, which is given by
\begin{equation}\label{eq:5}
    Q_s = \frac{2\pi f_0 L}{R_s},
\end{equation}
where $R_s$ is the equivalent series resistance (ESR), and L is the resonator inductance. 

\begin{figure*}[t]
    \centering
    \includegraphics[width=0.95\linewidth]{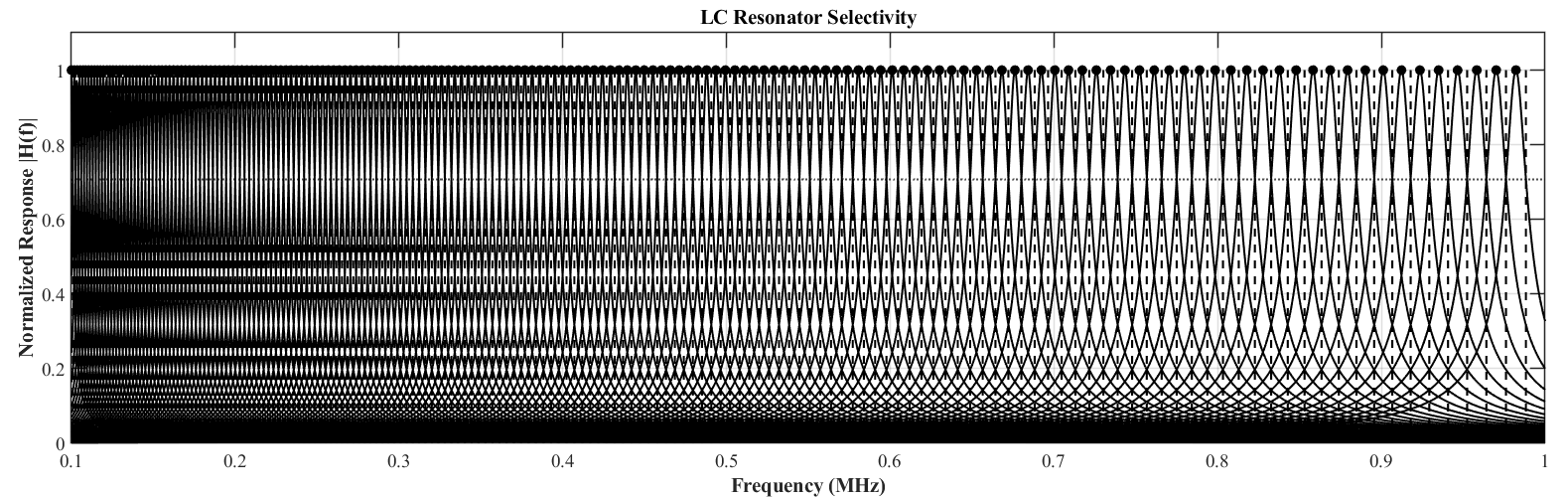}
    \vspace{-10px}
    \caption{Normalized frequency responses of 177 resonators designed with a constant inductor value. The plot shows the resonator magnitude responses across frequency, normalized to each resonator’s peak. The horizontal dashed line indicates the \SI{-3}{\decibel} threshold used to define the half-power bandwidth and assess spectral overlap.}
    \label{fig:3}
    \vspace{-15px}
\end{figure*}

As the inductive and capacitive reactances are equal in magnitude at resonance, the resonator Q-factor is primarily limited by dissipative losses and any additional loading from connected circuitry. For a practical LC tank, the loss can be approximated using a series-loss model as the sum of the dominant series contributions~\cite{Bahl2003LumpedElements}
\begin{equation}\label{eq:Rs_sum}
R_s(f_0) \approx R_L(f_0) + ESR_C(f_0) + R_{\mathrm{pcb}}(f_0),
\end{equation}
where $R_L$ denotes the inductor series loss, $ESR_C$ is the capacitor equivalent series resistance, and $R_{\mathrm{pcb}}$ accounts for interconnect losses in traces, vias, and contacts. In practice, capacitors typically exhibit low $ESR_C$, whereas inductors often dominate $R_s$ due to frequency-dependent winding and core losses, including skin effect, proximity effect, and ohmic loss~\cite{Kyaw2018}. 

Therefore, $\Delta f$ determines the approximate maximum number of individually addressable LC resonators, and subsequently robots, within a given frequency band $(f_{\min}, f_{\max})$. Since resonance frequencies are discretely allocated and $Q$ varies with frequency, we evaluate the achievable count using a discrete spacing rule. Let $f_i$ denote the center frequency of the $i$th resonator. To avoid significant overlap between adjacent resonators, the spacing between neighboring center frequencies is required to exceed the sum of their half-bandwidths with an additional guard band $e_f$,
\begin{equation}\label{eq:6}
f_{i+1}-f_i \ge \frac{\Delta f(f_i)}{2}+\frac{\Delta f(f_{i+1})}{2}+e_f.
\end{equation}
The number of addressable resonators is then obtained as
\begin{equation}\label{eq:7}
N=\max\{\,i\mid f_i\le f_{\max}\,\}\qquad f_1=f_{\min}.
\end{equation}

Array design can be carried out computationally by applying (\ref{eq:6}) across $(f_{\min},f_{\max})$ with the frequency-dependent $\Delta f(f)$ obtained from the measured or datasheet-based $Q(f)$. Non-idealities can be incorporated through the guard band $e_f$ and through uncertainty in resonator loss. Specifically, when $Q(f)$ is computed from the series-loss model, we account for resistance tolerances by using a worst-case effective series resistance $R_s^{\mathrm{eff}}(f)=R_s(f)+e_s$, which yields
\begin{equation}\label{eq:8}
Q^{\mathrm{eff}}(f)=\frac{2\pi f L}{R_s(f)+e_s},
\vspace{0.5px}
\end{equation}
and consequently $\Delta f(f)$ used in (\ref{eq:6}) is evaluated using $Q^{\mathrm{eff}}(f)$. In this formulation, $e_f$ increases the required spectral separation between adjacent resonators, while $e_s$ reduces the effective $Q$ through increased loss, both of which reduce the achievable $N$.

\section{Computational Scalability Analysis of Frequency-Selective Resonator Networks}

\subsection{Scalability analysis with Q-Factor and $R_s$}
We analyzed the number of individually addressable resonators that can be allocated within a given spectrum over bandwidths ranging from \SI{100}{kHz} to \SI{1}{MHz}, based on the theory presented in Section~\ref{subsec:q_factor}. Different actuator behavior is evident in the difference between the lower-impedance transistor base-drive path compared with the rectifier (trigger versus charger), because the bipolar junction transistor (BJT) draws additional current from the trigger resonators, effectively increasing the loaded loss and reducing the Q-factor. In a multi-robot system, operating at a larger frequency inherently widens the resonator bandwidth, reducing the frequency bandwidth for selectivity for a given Q-factor. For example, consider a Q-factor of 50. At \SI{1}{MHz} and \SI{10}{MHz}, (\ref{eq:4}) yields the half-power bandwidths ($\Delta f$) of \SI{20}{\kilo\hertz} and \SI{200}{\kilo\hertz}, respectively. To keep Q-factor constant throughout a frequency band is inherently a difficult task especially in higher RF bands where impedance matching is unforgiving. We shift our focus to the behaviors of an inductor as a key element to magnetic resonance power reception for independent actuator systems by modeling to reflect the equation derived to give further design insights into the capability of our system to be scaled in a variety of ways.

\subsection{Designing around the inductor}
The inductance is an easy variable to control in this frequency range. With the inductor identical, the inductive properties will be congruent from resonator to resonator. All coupling coefficients, coil geometries, can be held constant allowing us to have more insight into the circuit behavior and utilization of resonators for actuation devices. However, the effective series resistance as a function of frequency is a complicated derivation. To accurately model the bandwidth space of each resonator, we chose a commercially available power inductor with an expected $Q(f)$ curve. 

\begin{figure}
    \centering
    \includegraphics[width=3.4in]{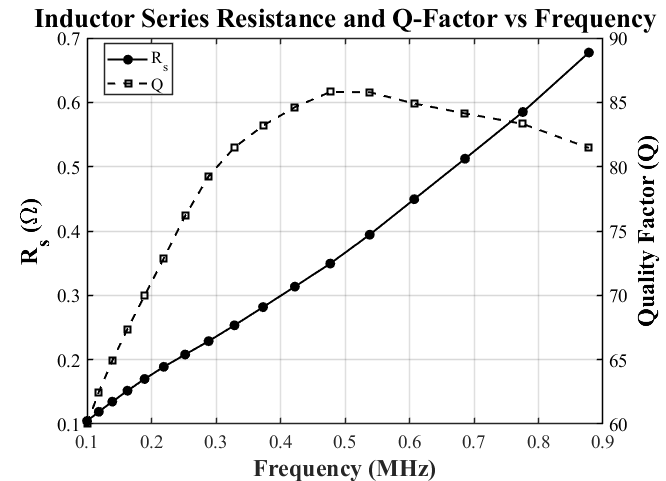}
    \vspace{-10px}
    \caption{Effective series resistance $R_s$ and corresponding Q-factor versus frequency for the selected inductor, showing frequency-dependent series damping of the LC resonator \cite{IHLP5050FD-AP}.}
    \label{fig:4}
    \vspace{-15pt}
\end{figure}

We developed a computational analysis to determine how many selectively addressable robots can be supported for a given set of component values. The results in Fig.~\ref{fig:3} show the normalized frequency responses of 177 resonators designed with a constant \SI{10}{\micro\henry} inductor, where the frequency-dependent loss $R_s(f)$ is incorporated through (\ref{eq:5}) using the datasheet-derived $Q(f)$ in \cite{IHLP5050FD-AP}. Each response is normalized to its peak, and the \SI{-3}{\decibel} line indicates the half-power threshold used to extract bandwidth and evaluate spectral overlap across the network. The Q-factor and $R_s$ as functions of frequency are shown in Fig.~\ref{fig:4}. With the inductor value constant, the capacitor values adjust to set the resonant frequency $f_0$ in (\ref{eq:1}) and the half-power bandwidth $\Delta f$ derived to fulfill it's portion of the spectrum and provide a numerical result to the number of resonators. 

The computational method provides a straightforward framework for designing large arrays of identical robots with selective operation. The number of resonators in the array scales with inductor size, as larger inductors provide stronger coupling to the transmitter, however it is evident that miniaturizing the robot with a smaller inductor increases the ratio of resistance to inductance which lowers the effective Q-factor. 

This finding makes miniaturization with smaller inductors more difficult because the effective Q-factor decreases, which broadens the resonator bandwidth and reduces the available spectral space for additional selectively addressable resonators. As the inductance is reduced further, the bandwidth continues to expand and frequency selectivity progressively degrades. The analysis in this section uses a fixed-inductor setup to isolate these spectral effects; the results suggest that maintaining selectivity at smaller scale will require mitigating loss and loading in the resonant network to preserve Q-factor.

\section{Centimeter-Scale RF-Triggered Actuator Fabrication}
\begin{table}[t]
\caption{Prototype Electronic Components and Specifications}
\label{tab:1}
\centering
\begin{tabular}{cc}
\toprule
\textbf{Designator} & \textbf{Specification} \\
\midrule
C1 & \SI{2}{\nano\farad} (C0G/NP0) \\
\midrule
L1, L2 & \SI{10}{\micro\henry} \\
\midrule
C2 & \SI{3.3}{\nano\farad}, \SI{3.9}{\nano\farad}, \SI{4.7}{\nano\farad} (C0G/NP0) \\
\midrule
C3 & \SI{1}{\micro\farad} \\
\midrule
D1, D2 & BAV21WSA \\
\midrule
C4 & \SI{330}{\micro\farad} \\
\midrule
D3 & 1N4749A \\
\midrule
Q1 & ZXTN19020CFFTA \\
\bottomrule
\end{tabular}
\vspace{-8px}
\end{table}

\subsection{Prototype design and component selection}
For rapid prototyping and validation of the proposed mechanism, we fabricated three centimeter-scale prototypes. Table~\ref{tab:1} presents the electronic components and their specifications for the prototypes. We chose the inductor values of \SI{10}{\micro\henry} to design around because of the relative feasibility to produce resonators with less than \SI{100}{\kilo\hertz} bandwidths while still maintaining strong coupling coefficients. To design and fabricate the three trigger LC resonators, we selected C0G/NP0 capacitors with values of \SI{3.9}{nF}, \SI{4.7}{nF}, and \SI{5.6}{nF} for the three devices, corresponding to designed trigger resonant frequencies of \SI{734}{kHz}, \SI{806}{kHz}, and \SI{876}{kHz}, respectively. The charging resonant frequency was set to \SI{1.125}{MHz} for all three devices using identical \SI{2}{nF} C0G/NP0 capacitors.

Each prototype has overall dimensions of \SI{25}{\milli\meter} $\times$ \SI{27}{\milli\meter} $\times$ \SI{13.5}{\milli\meter}. The rectifier diodes D1 and D2 are high-speed switching diodes (BAV21WSA). The DC-blocking capacitor C3 is \SI{1}{\micro\farad}, and the zener diode D3 has a breakdown voltage of \SI{24}{\volt} (1N4749A). The charge bank capacitor C4 is \SI{330}{\micro\farad}, and transistor Q1 (ZXTN19020CFFTA) has $\beta~\approx~300$ and a transition frequency of \SI{150}{\mega\hertz}. The rectifier diodes were selected to minimize forward loss and reduce power consumption. The zener diodes were used to help maintain energy circulation in the LC resonator promoting a prevention of unintended triggering.

\begin{table*}[h]
\vspace{8pt}
\caption{In Circuit: LC Resonator Measured Parameters}
\vspace{-10px}
\label{table2}
\begin{center}
\begin{tabular}{ccccccccc}
\toprule
  & L~\SI{}{(\micro\henry)} & C~\SI{}{(\nano\farad)} & Measured $\Delta f$\SI{}{(\kilo\hertz)} & Expected $f_0$ & Measured $f_0$ & Q-Factor & $E~at~f_0$~(\SI{}{\nano\joule}) \\
\midrule
\makecell{Charging LC1} & 10.0 & 2.0 & $\approx$20 & 1.125 MHz & 1.140 MHz & 57.00 & 14.89  \\
\midrule
\makecell{Charging LC2} & 10.0 & 2.0 & $\approx$20 & 1.125 MHz & 1.104 MHz & 55.20 & 14.06 \\
\midrule
\makecell{Charging LC3} & 10.0 & 2.0 & $\approx$20 & 1.125 MHz & 1.120 MHz & 56.00 & 14.44 \\
\midrule
\makecell{Trigger LC1} & 10.0 & 4.7 & $\approx$60 & 734 kHz & 734 kHz & 12.23 & 5.29 \\
\midrule
\makecell{Trigger LC2} & 10.0 & 3.9 & $\approx$55 & 806 kHz & 785 kHz & 14.27 & 4.39\\
\midrule
\makecell{Trigger LC3} & 10.0 & 3.3 & $\approx$60 & 876 kHz & 855 kHz & 14.25 & 3.81 \\
\bottomrule
\end{tabular}
\end{center}
\vspace{-20px}
\end{table*}

\subsection{Characterization of LC resonators} \label{subsec:LC-resonators}
\begin{figure}
\begin{tikzpicture}
\begin{axis}[
    width=3.42in,
    height=3.42in,
    title={In Circuit: LC Resonator Measured Frequency Response},
    xlabel={$Frequency (kHz)$},
    ylabel={$V_{\mathrm{rms}}$ (V)},
    xmin=600,
    xmax=1200,
    ymin=0,
    grid=both,
    legend pos= north west,
]
\addplot[
    red,
    thick,
    dashed,
    mark size=0.75pt,
    error bars/.cd,
    y dir=both,
    y explicit, 
    error bar style ={solid},
    error mark options = {rotate=90,black,mark size=3pt, }]
table[
    x=Freq,
    y=Vrms734,
    y error=Err734] {fig/LC_Data.txt};
\addlegendentry{Trigger 1 ~734 kHz}
\addplot[
    black,
    thick,
    dash dot,
    mark size=0.75pt,
    error bars/.cd,
    y dir=both,
    y explicit, 
    error bar style ={solid},
    error mark options = {rotate=90,black,mark size=3pt, }]
table[
    x=Freq,
    y=Vrms785,
    y error=Err785] {fig/LC_Data.txt};
\addlegendentry{Trigger 2 ~785 kHz}
\addplot[
    blue,
    thick,
    dotted,
    mark size=0.75pt,
    error bars/.cd,
    y dir=both,
    y explicit, 
    error bar style ={solid},
    error mark options = {rotate=90,black,mark size=3pt }]
table[
    x=Freq,
    y=Vrms855,
    y error=Err855] {fig/LC_Data.txt};
\addlegendentry{Trigger 3 ~855 kHz}
\addplot[
    red,
    thick,
    mark=*,
    mark size=0.75pt,
    error bars/.cd,
    y dir=both,
    y explicit, 
    error mark options = {rotate=90,black,mark size=3pt, }]
table[
    x=Freq,
    y=VrmsCH1,
    y error=ErrCH1] {fig/LC_Data.txt};
\addlegendentry{Charger 1 ~1.140 MHz}
\addplot[
    black,
    thick,
    mark=square*,
    mark size=0.75pt,
    error bars/.cd,
    y dir=both,
    y explicit, 
    error mark options = {rotate=90,black,mark size=3pt, }]
table[
    x=Freq,
    y=VrmsCH2,
    y error=ErrCH2] {fig/LC_Data.txt};
\addlegendentry{Charger 2 ~1.104 MHz}
\addplot[
    blue,
    thick,
    mark=triangle*,
    mark size=0.75pt,
    error bars/.cd,
    y dir=both,
    y explicit, 
    error mark options = {rotate=90,blue,mark size=3pt, }]
table[
    x=Freq,
    y=VrmsCH3,
    y error=ErrCH3] {fig/LC_Data.txt};
\addlegendentry{Charger 3 ~1.120 MHz}
\end{axis}
\end{tikzpicture}
\vspace{-10px}
    \caption{The red, black, and blue lines represent devices 1, 2, and 3 respectively as stated in the legend. The charging LCs are difficult to distinguish due to their close frequency overlap. The trigger LC resonators exhibit significantly lower Q-factor than the charging  LC resonators because the NPN transistor base current places a resistive load on the resonator, increasing energy loss and reducing effective Q. Error bars represent standard deviation (n=3).}
    \label{fig:2}
    \vspace{-12px}
\end{figure}
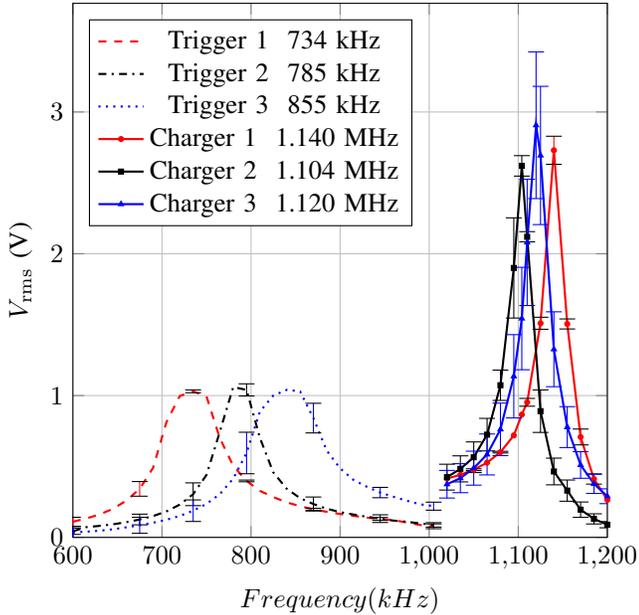

We characterized each of the three prototypes' two LC resonators to experimentally find their resonant frequency, bandwidth, Q-factor, and circulating energy, put together in Table~\ref{table2}. Each device was placed within the radius of the the transmitter on the same plane, and the root-mean-square voltage across each resonator was measured with an oscilloscope. The trigger LC circuits were measured from \SI{600}{\kilo\hertz} to \SI{1.005}{\mega\hertz} in \SI{15}{\kilo\hertz} intervals and using the same intervals for the charger LCs, from \SI{1.005}{\mega\hertz} to \SI{1.200}{\mega\hertz}. Measurements from three different positions within the transmitter plane were taken and averaged. The standard deviations (n=3) are shown as error bars in Fig. \ref{fig:2}.

Deviations from expected resonant frequency in most of the resonators are likely due to component tolerances (inductors $\pm20{\%}$ and capacitors $\pm20{\%}$) and stray reactive parasitics on the prototype board shifting $f_0$ away from the target value.

The Q-factors of the charging and trigger resonators are substantially different from $Q\approx55$ for the rectifier topology and  $Q\approx13$ for the transistor gate operation circuit. Two non-linear resistive loading effects are effectively causing energy loss from the resonator tank defined by parrallel resistance adding to the effective $R_s$ and influencing $Q^\text{eff}$ to decrease. The discrepency is due to once C4 stores charge it pulled less energy from the resonator meanwhile as long as the trigger LC is operational, the transistor gate will be operating and using energy. 

Each of the measured bandwidths are interpolated between the \SI{15}{kHz} measurement points, approximately \SI{20}{kHz} for charging resonators and \SI{60}{kHz} for the triggering resonators were accounted for. 

The peak circulating energy in each resonator was estimated by using (\ref{eq:2}) and oscilloscope measurements of the peak voltage across the LC while resonating. This value represents the instantaneous energy stored in the capacitor of the LC tank at the resonant peak, providing a single bound measurement of the total circulating energy. The triggers show smaller energy results resulting from lower Q values. These energy levels however confirm that sufficient energy is circulating to saturate the transistor briefly during cycles to discharge C4. 

\subsection{Electromagnetic actuation of a cantilever beam}

We integrated an onboard EM coil (L3), as shown in Fig.~\ref{fig:1}(a), and a magnetic cantilever beam; when the trigger LC drives the transistor gate, the coil actuates the beam. The EM coil was wound with AWG 38 copper wire around an iron core (Vacoflux 50, maximum permeability $\mu_{\max}\approx 7{,}000$) with a height of \SI{7}{\milli\meter} and a diameter of \SI{2}{\milli\meter}; the coil’s outer diameter was \SI{4.6}{\milli\meter}. A PCB-mounted cantilever beam with an integrated miniature permanent magnet (D11-N52, K\&J Magnetics, Inc.) was positioned above the EM coil. The beam was 3D-printed (Form 3, Formlabs Inc.) using Elastic 50A Resin.

\subsection{Transmitter coil}
The transmitter coil has a \SI{100}{\milli\meter} diameter, \SI{1}{\milli\meter} wire diameter, and 7 turns, with a DC resistance of \SI{0.3}{\ohm}. Its inductance was measured as \SI{10.5418}{\micro\henry} with a Q-factor of \SI{32.85}{} at \SI{300}{\kilo\hertz} (E4980AL, Keysight Technologies, Inc.). A function generator (DG1022, Rigol Technologies, Inc.) drives the coil through a simple matching network: a \SI{40}{\ohm} series resistor and a \SI{3.9}{\nano\farad} tuning capacitor. The capacitor value matches Device 2’s trigger capacitor (expected $f_0=\SI{806}{kHz}$) to center the excitation around the trigger frequencies, while the series resistor intentionally lowers the effective Q-factor to provide a broadband response spanning the receiver resonators’ operating range.

\section{Experimental Results}

\begin{figure}
    \centering
    \vspace{5pt}
    \includegraphics[width=0.95\linewidth]{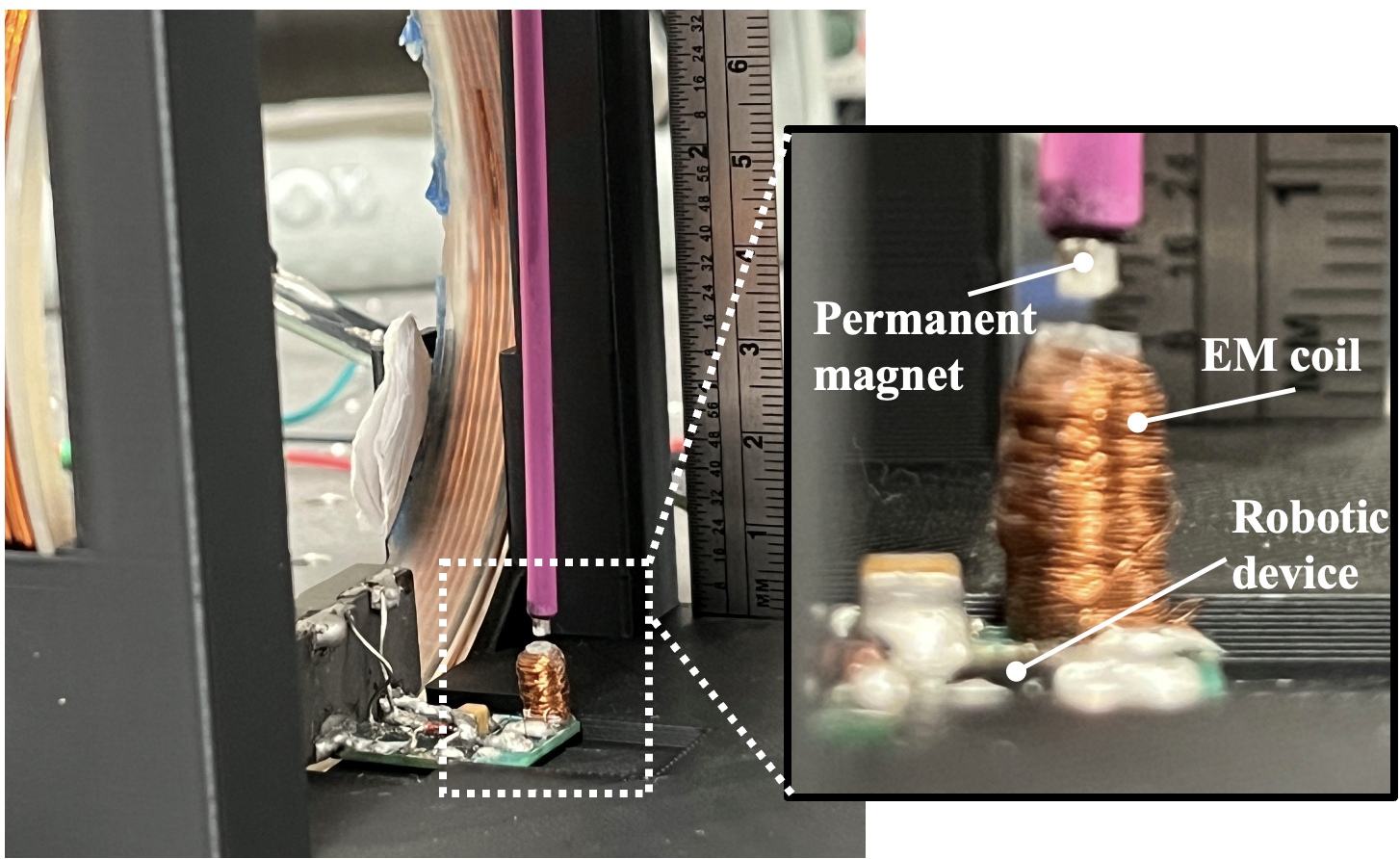}
    \vspace{-5px}
    \caption{Experimental setup of a 3D-printed configuring housing the GS0-30 load gram sensor, the transmitter coil, and each prototype as a fixed distance for actuation force characterization.}
    \label{fig:5}
    \vspace{-10px}
\end{figure}
This prototype demonstrates the ability to charge in one second and discharge to produce a significant magnetic force thus exhibiting the creation of an untethered pulsed electromagnetic actuator on the centimeter-scale. The actuator is designed with cantilever arm being pulled down by the coil under triggering frequencies. The resulting mechanical deflection confirms the actuator's effectiveness for latch-triggering applications.

\subsection{Electromagnetic actuation force characterization}
Fig.~\ref{fig:5} shows the setup used to characterize the EM actuation force, where a permanent magnet (D11-N52, K\&J Magnetics, Inc.) mounted on a rod is connected to a load cell (GSO-30, Transducer Techniques, LLC). In the robot units, the PCB-mounted cantilever beams serve as the actuators and exhibit visibly observable deflection upon triggering. Using the resonant charging and triggering frequencies for each device, the EM force was generated and measured. Measurements were taken with a gap distance of \SI{1}{mm} between the top surface of the EM coil and the permanent magnet. 

\begin{figure}[t]
    \centering
    \includegraphics[width=1\linewidth]{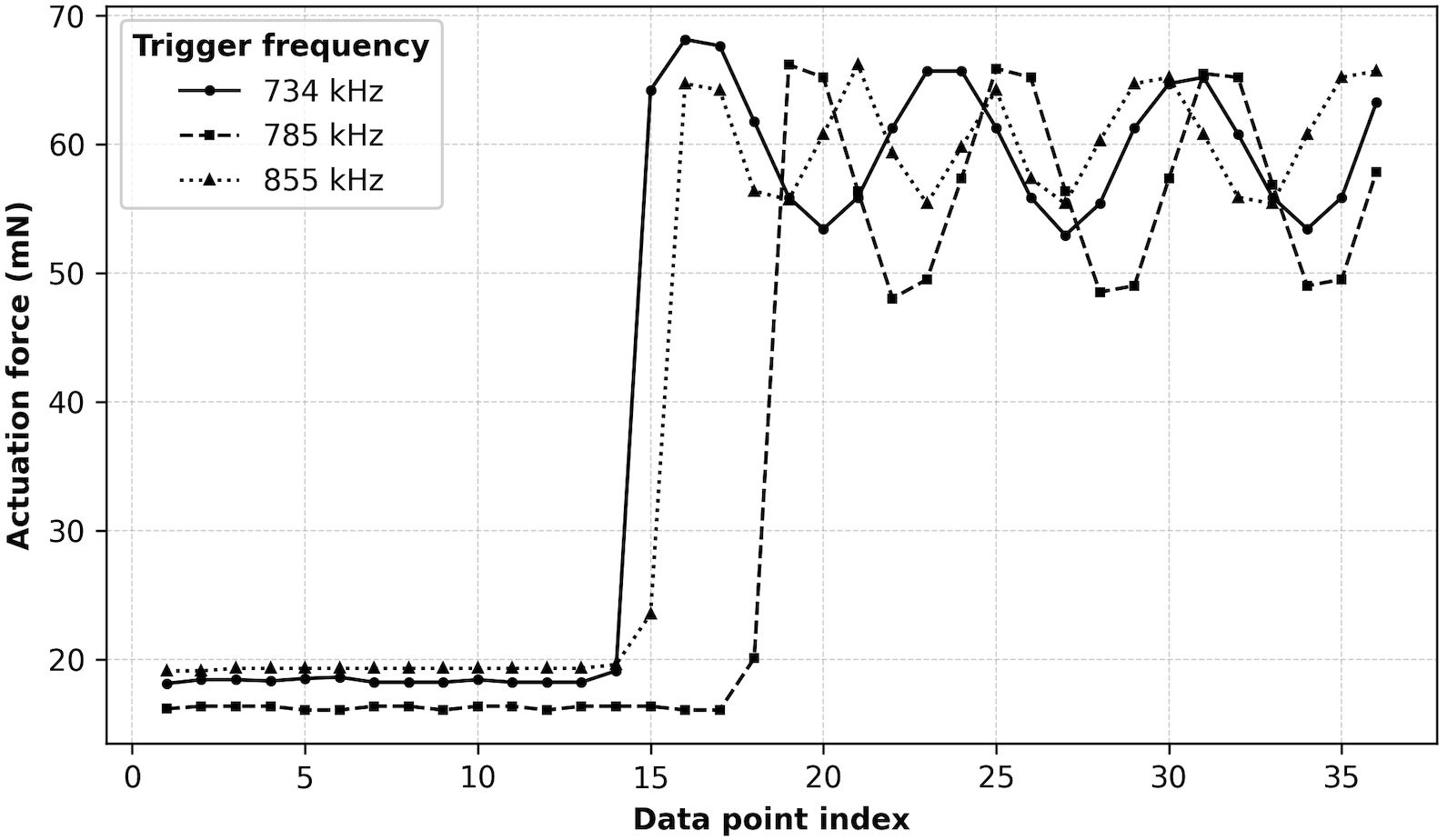}
    \caption{EM force measurement for three devices. Each device was set off a function generator to charge for one second and switch to the trigger frequency for one second.}
    \label{fig:6}
      \vspace{-10px}
\end{figure}

Fig.~\ref{fig:6} shows the force measurements of the three devices. The peaks of each reading were above \SI{60}{mN} and below \SI{70}{mN}. The force traces exhibit noticeable point-to-point fluctuations rather than a smooth plateau. To investigate this behavior, we probed the discharge of C4 (charge bank) using an oscilloscope (DS1102, Rigol Technologies, Inc.). The measurements revealed microsecond-scale voltage oscillations that can repeatedly cross the gate threshold and re-open the transistor, producing multiple closely spaced actuation pulses while progressively discharging C4; this high-speed retriggering is not apparent from visual observation alone. To mitigate this effect, we will add gate damping to suppress ringing and prevent unintended retriggering. To increase the peak actuation force without extending the cycle time, we will boost the charge rate and charge-bank power by using a higher-current charging stage and a low-loss, low-ESR charge/discharge path to deliver a larger current pulse within the same time window.

\subsection{Selective triggering validation}
To experimentally determine the frequency bandwidth over which a single excitation could trigger more than one actuator, we swept the input frequency around each device’s designed resonant frequency and monitored the discharge behavior of the charge bank capacitor (C4). We used the onset of C4 discharge as the triggering criterion to quantify each device’s effective trigger bandwidth, and then analyzed the overlap in triggering frequencies across the three devices.

\begin{figure}
    \centering
    \includegraphics[width=1\linewidth]{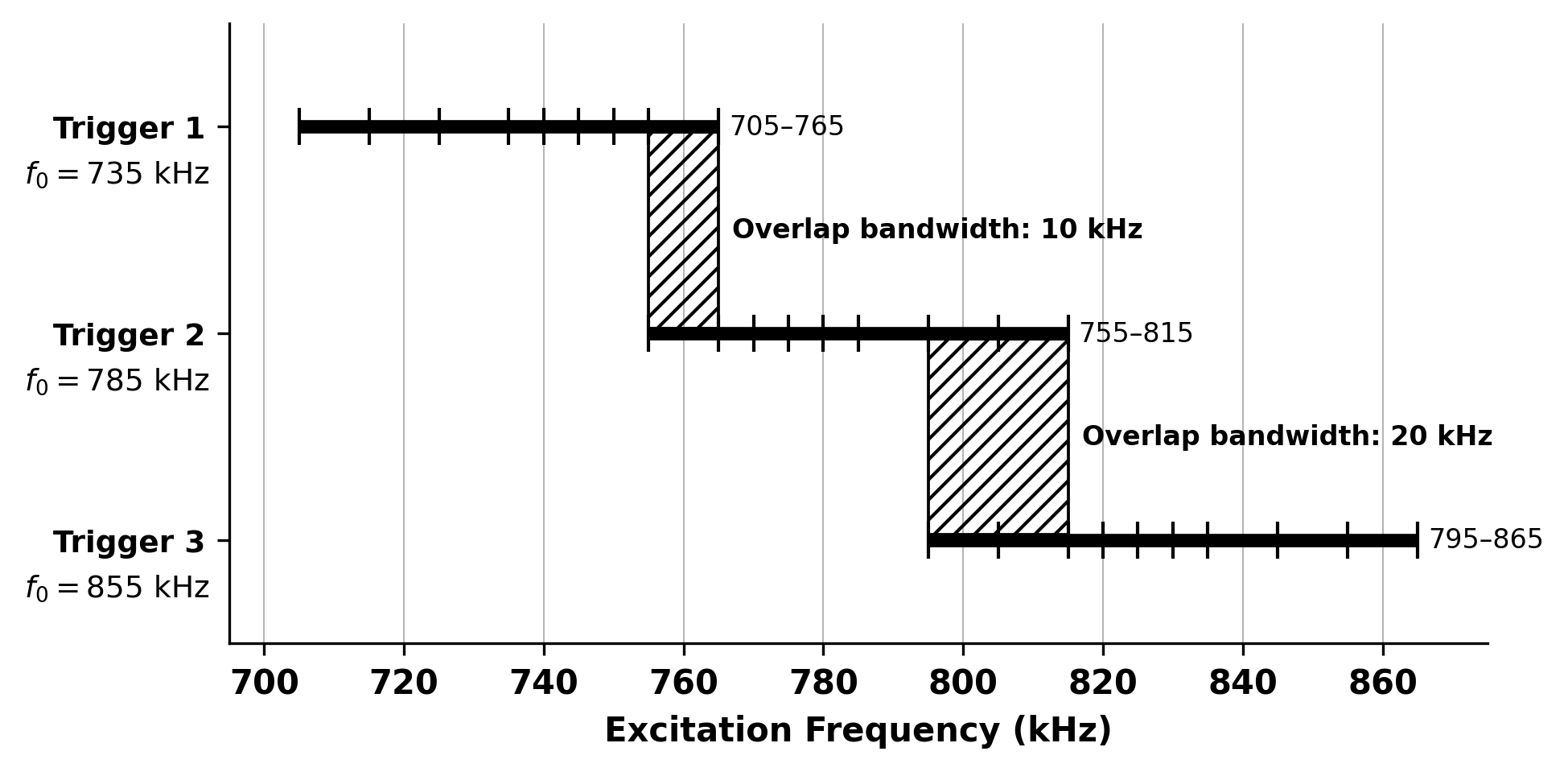}
    \caption{Selective activation frequency bands for three trigger LC resonators. Solid bars indicate the measured excitation frequency ranges that trigger each device, and tick marks denote the tested frequencies. Hatched regions highlight the overlap between adjacent triggers (10 kHz for Triggers 1–2 and 20 kHz for Triggers 2–3).}
    \label{fig:7}
    \vspace{-15px}
\end{figure}

Fig.~\ref{fig:7} shows the measured triggerable frequency bands for the three devices and highlights the overlap regions. Trigger 1 \((f_0=\SI{735}{kHz})\) was activated over \SI{705}{kHz}$\sim$\SI{765}{kHz}, Trigger 2 \((f_0=\SI{785}{kHz})\) over \SI{755}{kHz}$\sim$\SI{815}{kHz}, and Trigger 3 \((f_0=\SI{855}{kHz})\) over \SI{795}{kHz}$\sim$\SI{865}{kHz}. The overlap between Triggers 1 and 2 is limited to \SI{10}{kHz} (\SI{755}{kHz}–\SI{765}{kHz}), and the overlap between Triggers 2 and 3 is \SI{20}{kHz} (\SI{795}{kHz}–\SI{815}{kHz}). These overlap regions reflect the finite resonator bandwidth governed by the Q-factor, as given by (\ref{eq:4}); increasing Q reduces the bandwidth, thereby reducing spectral overlap and lowering the likelihood of cross-triggering, consistent with the loss mechanisms discussed earlier.

\begin{figure*}[t]
    \centering 
    \includegraphics[width = 1\linewidth]{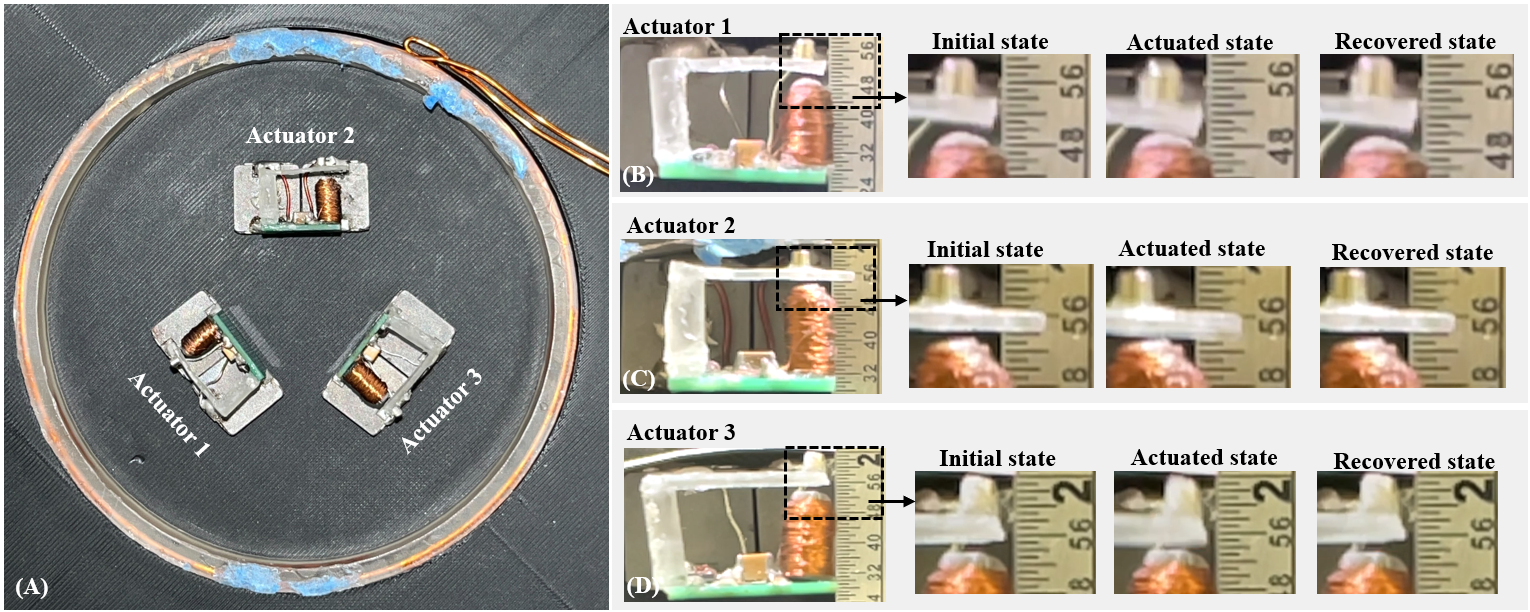}
    \caption{Cantilever beam deflections with one-second charge and one-second discharge cycles. (A) Experimental configuration showing the transmitter coil and three actuators positioned near the transmitter and aligned parallel to the coupling axis. (B)-(D) Time-lapse sequences for each actuator showing the initial state, actuated state, and recovered state, demonstrating selective actuation of individual units.}
    \label{fig:8}
    \vspace{-8px}
\end{figure*}

\subsection{Selective actuation of an integrated mechanical beam}
To demonstrate selective actuation of the integrated mechanical beam as a functional unit, Actuators 1, 2, and 3 were arranged in a circular configuration around the transmitter coil, as shown in Fig.~\ref{fig:8}(a). The transmitter was driven by one second charge and one second discharge times of the resonant charging and triggering frequencies in each device mentioned in Table~\ref{table2}. Fig.~\ref{fig:8}(b), (c), and (d) show deflection of Actuators 1, 2, and 3 captured in a slow-motion video at 240 FPS next to a US Standard System ruler containing 1/64th inch graduations. The visible deflection of about one graduation in each device likely does not capture the true displacement. However, the visual transition between initial, actuated, and recovered states in all three devices demonstrates the capacity of the magnetic force to create flexion in the structure and the elasticity of the structure to recover. Each actuator exhibits a complete latching cycle, deflection under actuation followed by elastic recovery, confirming that the generated magnetic field force exceeds cantilever restoring force threshold. The three devices actuate independently with no observable cross-triggering, consistent to the selectivity measured in Fig. \ref{fig:7} further showing frequency selectivity. The orientation of the devices demonstrates that the proximal arrangement has no negative impact on the individuality of each device.

\section{Discussion}
A primary limitation of the current prototypes is resistive loss in the actuation and triggering pathways. In future PCB designs, lower-resistance actuation paths could be achieved by selecting power transistors with low $R_{ds}(\mathrm{on})$ in the on state. The force output in the current design was limited by the NPN switching characteristics, which introduce additional resistance in the actuation path. Reducing the switch resistance would increase the peak current through L3 and thereby increase force output. The measured triggering Q of 12, compared with the charging Q of 57, reflects resistive loading at the NPN base junction. Replacing the NPN BJT with a high-impedance MOSFET would increase the Q seen by the trigger resonator by reducing loading, thereby lowering the energy threshold required for actuation. This improvement could also enable longer activation distances and make further miniaturization more feasible by reducing the harvested energy required for triggering.

Applying the derived scaling relationship to smaller inductor values shows that sustaining selectivity at reduced inductance primarily requires maintaining a high Q-factor. When the inductor is reduced from \SI{10}{\micro\henry} to \SI{1}{\micro\henry} within the same \SI{100}{\kilo\hertz} to \SI{1}{\mega\hertz} band, the predicted addressable count decreases because the effective resonator bandwidth increases as L decreases and resistive losses become more influential. This behavior points to two primary design factors for smaller-scale designs, reducing the effective series resistance and reducing loading from the surrounding circuitry. 

Based on these two factors, we identify several directions to improve selectivity at reduced scale. Microelectromechanical inductor fabrication techniques may provide higher inductance density at reduced physical scales, although integration with WPT energy harvesting circuits for robotic actuation remains to be demonstrated. In the current architecture, WPT must support both energy harvesting for actuation and frequency-selective triggering, which places competing demands on inductors, switching paths, and storage elements. Adding onboard energy storage would decouple these functions by allowing frequency selectivity to focus on triggering while actuation energy is supplied from stored charge. This shift can reduce the need for two large inductors, reduce PCB footprint, and improve actuation consistency that is currently sensitive to the fixed one-second charge and discharge timing.


\addtolength{\textheight}{-12cm}   


\bibliographystyle{IEEEtran}
\bibliography{ AIM}

@article{Boyvat_Koh_Wood_2017, 
title={Addressable wireless actuation for multijoint folding robots and Devices}, 
volume={2}, 
DOI={10.1126/scirobotics.aan1544}, 
number={8}, 
journal={Science Robotics}, 
author={Boyvat, Mustafa and Koh, Je-Sung and Wood, Robert J.}, 
year={2017}, 
month={Jul}
}

@article{Floyd_Diller_Pawashe_Sitti_2011, title={Control methodologies for a heterogeneous group of untethered magnetic micro-robots}, volume={30}, DOI={10.1177/0278364911399525}, number={13}, journal={The International Journal of Robotics Research}, author={Floyd, Steven and Diller, Eric and Pawashe, Chytra and Sitti, Metin}, year={2011}, month={Mar}, pages={1553–1565}}

@ARTICLE{6056575,
  author={Diller, Eric and Floyd, Steven and Pawashe, Chytra and Sitti, Metin},
  journal={IEEE Transactions on Robotics}, 
  title={Control of Multiple Heterogeneous Magnetic Microrobots in Two Dimensions on Nonspecialized Surfaces}, 
  year={2012},
  volume={28},
  number={1},
  pages={172-182},
  doi={10.1109/TRO.2011.2170330}}

@article{Tottori_Zhang_Peyer_Nelson_2013, title={Assembly, disassembly, and anomalous propulsion of microscopic helices}, volume={13}, DOI={10.1021/nl402031t}, number={9}, journal={Nano Letters}, author={Tottori, Soichiro and Zhang, Li and Peyer, Kathrin E. and Nelson, Bradley J.}, year={2013}, month={Aug}, pages={4263–4268}}

@ARTICLE{7394119,
  author={Wong, Denise and Steager, Edward B. and Kumar, Vijay},
  journal={IEEE Robotics and Automation Letters}, 
  title={Independent Control of Identical Magnetic Robots in a Plane}, 
  year={2016},
  volume={1},
  number={1},
  pages={554-561},
  doi={10.1109/LRA.2016.2522999}}

@ARTICLE{8440784,
  author={Kantaros, Yiannis and Johnson, Benjamin V. and Chowdhury, Sagar and Cappelleri, David J. and Zavlanos, Michael M.},
  journal={IEEE Transactions on Robotics}, 
  title={Control of Magnetic Microrobot Teams for Temporal Micromanipulation Tasks}, 
  year={2018},
  volume={34},
  number={6},
  pages={1472-1489},
  doi={10.1109/TRO.2018.2861901}}

@INPROCEEDINGS{525362,
  author={Inoue, T. and Iwatani, K. and Shimoyama, I. and Miura, H.},
  booktitle={Proceedings of 1995 IEEE International Conference on Robotics and Automation}, 
  title={Micromanipulation using magnetic field}, 
  year={1995},
  volume={1},
  number={},
  pages={679-684 vol.1},
  doi={10.1109/ROBOT.1995.525362}}

@article{Chowdhury_Jing_Cappelleri_2015, title={Towards independent control of multiple magnetic mobile Microrobots}, volume={7}, DOI={10.3390/mi7010003}, number={1}, journal={Micromachines}, author={Chowdhury, Sagar and Jing, Wuming and Cappelleri, David}, year={2015}, month={Dec}, pages={3}}

@article{Amoudruz_Koumoutsakos_2021, title={Independent Control and path planning of microswimmers with a uniform magnetic field}, volume={4}, DOI={10.1002/aisy.202100183}, number={3}, journal={Advanced Intelligent Systems}, author={Amoudruz, Lucas and Koumoutsakos, Petros}, year={2021}, month={Dec}}

@article{Ishiyama_Sendoh_Arai_2002, title={Magnetic micromachines for medical applications}, volume={242–245}, DOI={10.1016/s0304-8853(01)01181-7}, journal={Journal of Magnetism and Magnetic Materials}, author={Ishiyama, K. and Sendoh, M. and Arai, K.I.}, year={2002}, month={Apr}, pages={41–46}}

@ARTICLE{8361455,
  author={Narayanamoorthi, R. and Juliet, A. Vimala and Chokkalingam, Bharatiraja},
  journal={IEEE Sensors Journal}, 
  title={Frequency Splitting-Based Wireless Power Transfer and Simultaneous Propulsion Generation to Multiple Micro-Robots}, 
  year={2018},
  volume={18},
  number={13},
  pages={5566-5575},
  doi={10.1109/JSEN.2018.2838671}}

@ARTICLE{Kyaw2018,
  author={Kyaw, Phyo Aung and Stein, Aaron L. F. and Sullivan, Charles R.},
  journal={IEEE Transactions on Power Electronics}, 
  title={Fundamental Examination of Multiple Potential Passive Component Technologies for Future Power Electronics}, 
  year={2018},
  volume={33},
  number={12},
  pages={10708-10722},
  doi={10.1109/TPEL.2017.2776609}}

@article{Coakley2003QEstimation,
  author  = {Coakley, Kevin J. and Splett, Jolene D. and Janezic, Michael D. and Kaiser, Raian F.},
  title   = {Estimation of {Q}-Factors and Resonant Frequencies},
  journal = {IEEE Transactions on Microwave Theory and Techniques},
  year    = {2003},
  volume  = {51},
  number  = {3},
  pages   = {862--868},
  doi     = {10.1109/TMTT.2003.808578},
}

@article{SongNatCommun2025RFSelectiveLCE,
  author  = {Song, Yiwen and Li, Zefang and Zadan, Mason and Wang, Jingxian and Kumar, Swarun and Majidi, Carmel},
  title   = {Frequency-selective actuation of liquid crystalline elastomer actuators with radio-frequency},
  journal = {Nature Communications},
  year    = {2025},
  volume  = {16},
  number  = {1},
  pages   = {7292},
  doi     = {10.1038/s41467-025-62313-9},
}

@article{RichterAdvs2023LocallyAddressable,
  author  = {Richter, Michiel and Sikorski, Jakub and Makushko, Pavlo and Zabila, Yevhen and Venkiteswaran, Venkatasubramanian Kalpathy and Makarov, Denys and Misra, Sarthak},
  title   = {Locally Addressable Energy Efficient Actuation of Magnetic Soft Actuator Array Systems},
  journal = {Advanced Science},
  year    = {2023},
  volume  = {10},
  number  = {24},
  pages   = {e2302077},
  doi     = {10.1002/advs.202302077},
}

@article{TakeuchiSensActA2002SelectiveDrive,
  author  = {Takeuchi, S. and Futai, N. and Shimoyama, I.},
  title   = {Selective Drive of Electrostatic Actuators Using Remote Inductive Powering},
  journal = {Sensors and Actuators A: Physical},
  year    = {2002},
  volume  = {95},
  number  = {2-3},
  pages   = {269--273},
  doi     = {10.1016/S0924-4247(01)00738-5},
}

@article{MateenMicrosystNanoeng2016WirelessActuation,
  author  = {Mateen, Farrukh and Maedler, Carsten and Erramilli, Shyamsunder and Mohanty, Pritiraj},
  title   = {Wireless actuation of micromechanical resonators},
  journal = {Microsystems \& Nanoengineering},
  year    = {2016},
  volume  = {2},
  pages   = {16036},
  doi     = {10.1038/micronano.2016.36},
}

@book{Bahl2003LumpedElements,
  author    = {Bahl, Inder J.},
  title     = {Lumped Elements for RF and Microwave Circuits},
  publisher = {Artech House},
  year      = {2003},
  isbn      = {9781580533096},
  note      = {See Ch.~1 (lumped modeling of interconnects), Ch.~3 Sec.~3.1.1 (inductor conductor loss), and Ch.~5 Sec.~5.2.6 (capacitor ESR).},
}

@Manual{IHLP5050FD-AP,
  title        = {IHLP® Automotive Inductors, Low AC Loss, High Temperature (155 °C) Series},
  year         = {2026},
  number       = {34569},
  note         = {Rev. 16-Feb-2026},
  organization = {Vishay Dale},
  url          = {https://www.vishay.com/en/product/34569/},
}
\end{document}